\documentclass[letterpaper, 10 pt, conference, dvipsnames]{ieeeconf}  
\usepackage{amsmath}
\usepackage{graphicx}
\usepackage{multirow}
\graphicspath{ {Figures/} }
\usepackage{subcaption}
\usepackage{tikz}
\usepackage{mdframed}

\DeclareRobustCommand{\legendsquare}[1]{%
  \tikz[baseline=(a.south)]{\node[#1, inner sep=.8ex, outer sep=0] (a) {};}%
}

\IEEEoverridecommandlockouts                              

\title{\LARGE \bf
Siamese Networks for Weakly Supervised Human Activity Recognition
}

\author{Taoran Sheng$^{1}$ and Manfred Huber$^{1}$
\thanks{$^{1}$ Taoran Sheng and Manfred Huber are with the Department of Computer Science and Engineering, University of Texas at Arlington, Arlington, TX 76019-0015, USA 
        {\tt\small taoran.sheng@mavs.uta.edu\qquad  huber@cse.uta.edu}}%
}

\usepackage{mycomments}
\begin{document}

\maketitle
\thispagestyle{empty}
\pagestyle{empty}

\begin{abstract}

Deep learning has been successfully applied to human activity recognition. However, training deep neural networks requires explicit\add{ly} labeled data which is difficult to acquire. In this paper, we present a model with multiple siamese networks that are trained by using only the information about the similarity between pairs of data samples without knowing the explicit labels. The trained model maps the activity data sample\add{s} into fixed size representation vectors such that the distance between the vectors in the representation space approximates the similarity of the data samples in the input space. Thus, the trained model can work as a metric for a wide range of different clustering algorithms. The training process minimizes a similarity loss function that forces the distance metric to be small for pairs of samples from the same kind of activity, and large for pairs of samples from different kinds of activities. We evaluate the model on three datasets to verify its effectiveness in segmentation and recognition of continuous human activity sequences.

\end{abstract}

\section{INTRODUCTION}

Human Activity Recognition (HAR) is critical in many research areas\del{:}\add{, including} human computer interaction, smart assistive technologies etc. These areas use HAR system\add{s} to provide information about people's activities and behaviors. The common way to implement such a system is \add{by} coll\del{o}\add{e}cting data from \add{environmental or} wearable sensors and processing \add{this} data with machine learning algorithms. 
One \del{kind} of the dominant framework\del{s} \add{types} \cite{transitionAware} used in HAR system\add{s} relies on $(i)$ sliding window segmentation of time series data recorded by wearable sensors, $(ii)$ manually designed features, such as statistical mean, variance, entropy of the signal, \add{and} features extracted from the frequency domain, \add{and} $(iii)$ different supervised classification algorithms to recognize activity. This \del{kind}\add{type} of framework\del{s} perform\add{s} well; however, domain knowledge is required to determine the window size, sliding speed, and to design the features manually. 
Recently, many deep neural networks (DNN) \del{are}\add{have been} developed and applied in \del{the} HAR system\add{s} \del{\cite{cnnHAR} \cite{dnnHARbenchmark} \cite{DeepCNNLSTMSensor}}\add{\cite{cnnHAR,dnnHARbenchmark,DeepCNNLSTMSensor}}. These methods can automatically extract features from the data without any domain knowledge and \del{are proved}\add{have been shown} to be useful in HAR applications. But, they still require explicit labels to supervise the training of the model and \add{usually a} hand designed sliding window method to segment the time series data.
Our work is motivated by two factors: enabling automatic\del{ally} segmentation of the time series \mdel{data} and limit\add{ing} the supervision \add{required while}\del{on} learning a recognition model. In this paper\mdel{,} we propose an approach based on multiple siamese networks to segment and recognize human activities in sensor data stream\add{s}. The proposed approach learns one siamese network to automatically segment the times series \mdel{data} without \madd{manually} designing \mdel{the}\madd{a} sliding window\mdel{ manually}, and another siamese network to provide a similarity metric that can be used to cluster the activities without using \del{any}\add{explicitly} labeled data.

\section{RELATED WORK}
The typical process \cite{tutorial} of human activity recognition includes signal preprocessing, data segmentation, feature extraction, and applying machine learning to recognize \add{each} activity. In this work, we mainly focus on the data segmentation stage and the activity recognition stage.

\subsection{Time Series Data Segmentation}

The data segmentation stage identifies the segments in the data stream that are likely to be an activity and determines the start time $t_{start}$ and end time $t_{end}$ of the activity. Once all the activity segments in the data stream are identified, they can be fed into the recognition module. 

Segmenting a continuous sensor stream is difficult. Because different activities continuously performed by people smoothly blur into each other, \add{they are} not clearly separated by a predefined posture or \mdel{a} pause. \mdel{

}One common approach used to segment the sequence data is \add{using a} sliding window \del{\cite{window1} \cite{window2}}\add{\cite{window1,window2}}. A window with predefined \mdel{window} length and step size moves over the data stream, and the data sequence contained within the window is used as one data sample, which has one data label. This method, however, will introduce inaccuracy into the segmentation borders. \del{L}\add{A l}arger window size is needed to catch complex activit\del{y}\add{ies}\mdel{. The more complex the activity, the larger the window should \add{generally} be}; but the larger the window, the less accurately the segmentation borders can be defined. Another segmentation method is based on signal energy \cite{energysegment}. It assumes that the intensities of different activities are different and that the different intensities can be used as an indicator to determine the borders of an activity. \del{\add{Ashbrook and Starner} \cite{extrasensorsegment} proposed \add{a} method that segments one sensor data stream using additional sensor reading\add{s}.}

\subsection{Feature Extraction and Recognition}

In order to recognize \add{an} activity, discriminative features are needed. They can be designed with domain knowledge manually or learned by neural networks automatically. 
Hand designed features are widely used in HAR \del{\cite{transitionAware} \cite{wisdm}}\add{\cite{transitionAware,wisdm}}. They include statistical features, such as mean, variance and entropy, or features extracted \del{from}\add{in the} frequency domain using a fourier transform, wavelet transform \cite{waveletFeature} or discrete cosine transform \cite{dctFeature} etc. The advantage of these features is that they can be derived from the signal easily and \del{are proved}\add{have been shown to be} effective in the HAR system.

With the advances of DNN, many HAR systems adopt DNN to allow automatic extraction of meaningful features. \add{Zeng et al.} \cite{cnnHAR} used convolutional neural networks (CNN) to capture\del{s} local dependencies and \add{identify}\del{the} scale invariant features of the activity signals. The local connectivity constraint between adjacent layers in \add{the} CNN forces the model to capture the local dependencies. \add{Morales and Roggen} \cite{cnnLSTMHAR} proposed models that combine CNN and Long Short Term Memory cells (LSTM) \cite{lstm1997}. LSTM keeps track of an internal state that represents its memory. The memory state eases the learning of long time scale temporal dependencies\del{,} \add{and} helps the HAR system to model more complex activities.

\subsection{Siamese Neural Networks}
A Siamese Network \cite{signature1993} is a neural network with two branches and tied weights. It processes two different inputs and yields two comparable representation vectors, which represent the features of each input\add{,} respectively\add{.}\del{,} \del{t}\add{T}hen the representation vectors are fed deeper into the network, which consists of a predefined metric layer \cite{sentenceSiamese} or a learned metric network to measure how similar the two inputs are. The siamese network is widely used to learn non-linear metrics, and has been successfully applied in computer vision, speech recognition\madd{,} etc. \madd{A s}\mdel{S}iamese CNN \mdel{can be}\madd{has been} used to learn \add{a} complex similarity metric for face verification \madd{in} \cite{faceverification}. \add{Mueller and Thyagarajan} \cite{sentenceSiamese} proposed a siamese RNN to measure semantic similarity between sentences. \add{Zeghidour et al.} \cite{jointlearning} used a multi-output triamese network to identify the speaker and phonetic similarities jointly. \add{Neculoiu et al.}  \cite{textSiamese} used a siamese RNN for job title normalization\mdel{, a common task in information extraction for recruitment and social network analysis}.

\section{PROPOSED APPROACH}

\begin{figure}[thpb]
      \centering
      \includegraphics[scale=0.2]{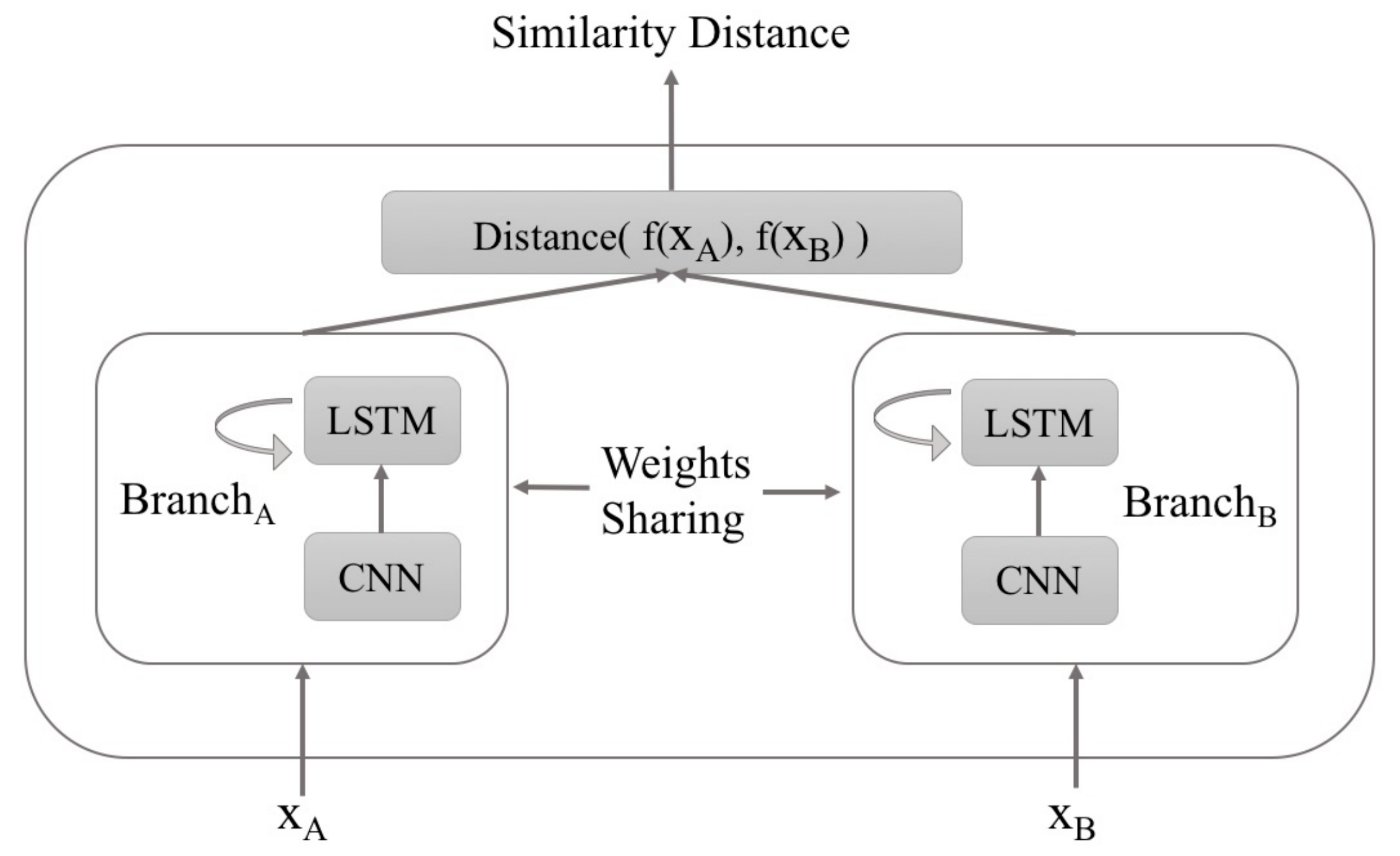}
      \caption{The basic siamese architecture used in our model}
      \label{siameseArch}
   \end{figure}

Our proposed model contains two modules\del{:}\add{, a} segmentation module and \add{a} recognition module. Both of them are using \del{the}\add{a} similar siamese architecture (see Fig.\add{~}\ref{siameseArch}) but with different LSTM structures and similarity functions.

\subsection{Siamese Architecture in the Proposed Approach} 
\add{\label{se:Siamese}}
As shown in Fig.\add{~}\ref{siameseArch}, given \del{the}\add{an} input pair of human activity data sequences $(x_{A}, x_{B})$, the siamese network learns to map \mdel{the input pair}\madd{it} to the representation space $(f(x_{A})$, $f(x_{B}))$ $\in$ $R^{d}$\mdel{, t}\madd{. T}hen the layers above the dual-branch represent a similarity function that measures the distance between these two representation vectors. We will detail the siamese architecture in this section. The segmentation module and recognition module will be introduced in \del{s}\add{S}ubsection\madd{s}\add{~\ref{se:SegmentationModule}}\del{ $B$} and\mdel{ \del{s}\add{S}ubsection}\add{~\ref{se:RecognitionModule}}\del{ $C$}\add{,} respectively. 

\begin{figure}[thpb]
      \centering
      \includegraphics[scale=0.14]{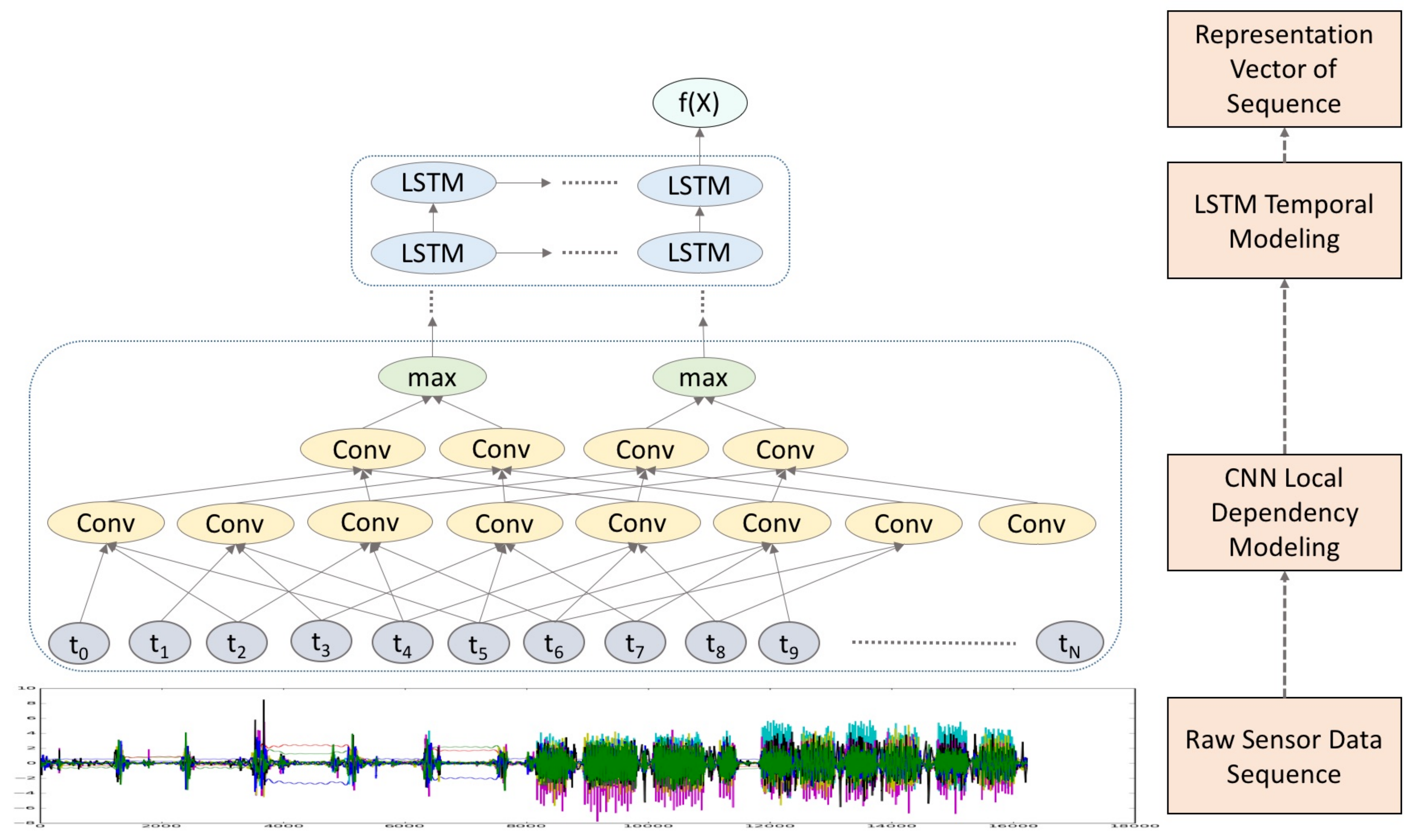}
      \caption{Details of a single branch in the siamese network}
      \label{singleBranch}
   \end{figure}

Our siamese networks share weights across \mdel{its}\madd{their} two branches. Each branch uses the same building blocks: $(i)$ Dilated temporal convolutional layer\madd{s}, \add{and} $(ii)$ Residual LSTM layer\madd{s}. \del{And f}\add{F}ully connected layers (FC) will be adopted to process the outputs from both branches. Fig.\add{~}\ref{singleBranch} illustrates details in each \mdel{single} branch.

\begin{itemize}

\item \add{\it Dilated Temporal Convolutional Layer}\del{Dilated Temporal Convolutional layer}

The temporal convolutional layer is applied to the raw data sequences with the aim of matching the local pattern of the input data and enabling translational invariance of each pattern in the activity data sequence. \del{And t}\add{T}o increase the receptive field of the temporal convolutional layer, we adopt dilated temporal convolution, as it supports faster expanding receptive fields without losing resolution or coverage \cite{MULTISCALCONtext}. Batch Normalization (BN) \cite{BNA} is applied after each temporal convolutional layer and before the non-linear activation function. BN can accelerate the learning process by preventing the internal covariate shift problem, \del{letting}\add{allowing} each intermediate layer \del{don't}\add{not to} have to adapt \del{themselves}\add{individually} to a new distribution in every training step. In addition, BN provides a slight regularization effect to the network\del{,} and \del{it} can also prevent the early saturation of \add{the} non-linear activation function. To further reduce the temporal dimensions of the data sequence\del{,} while introducing slight translational invariance in time, we insert a temporal max-pooling layer after every two dilated temporal convolutional layers. \del{The m}\add{M}ax-pooling outputs the maximum within the region of a predefined pooling size and corresponds to a subsampling. The output of the max-pooling layer is fed into the subsequent LSTM layers.

\item \add{\it Residual LSTM Layer}\del{Residual LSTM layer}

The LSTM layers are responsible for model\mdel{l}ing the higher level temporal patterns in the data. The definition of an LSTM cell is as follows:

~\vspace*{-0.25in}
\begin{align}
& i_t = \sigma_g(W_i*x_t + U_i*h_{t-1} + b_i) \label{lstmi}\\
& f_t = \sigma_g(W_f*x_t + U_f*h_{t-1} + b_f) \label{lstmf}\\
& \tilde{c_t} = \sigma_c(W_c*x_t + U_c*h_{t-1} + b_c) \label{lstmcc}\\
& c_t = f_t \circ c_{t-1} + i_t \circ \tilde{c_t} \label{lstmc}\\
& o_t = \sigma_g(W_o*x_t + U_o*h_{t-1} + b_o) \label{lstmo}\\
& h_t = o_t \circ \sigma_h(c_t) \label{lstmh}
\end{align}

\madd{A} LSTM \madd{unit} maintains a cell state $c_{t}$ \mdel{in its structure} which contains the information from the past. Eq.\add{~}\eqref{lstmi} defines an input gate, which decides how much \add{of} the input $x_{t}$ will be fed into the cell state. Eq.\add{~}\eqref{lstmf} defines a forget gate which decides \add{to what level}\del{what} information will be removed from the cell state $c_{t}$. Eq.\add{~}\eqref{lstmcc} defines the way to compute a cell state candidate $\tilde{c_{t}}$ that will be used to update the cell state. Then we can use the gates and the candidate to update the cell state as shown in Eq.\add{~}\eqref{lstmc}. Eq.\add{~}\eqref{lstmo} defines an output gate. The output will be based on the cell state\del{,} but will be a filtered version. \del{And t}\add{T}he output gate works as the filter to decide what parts of the cell state will be the output, as shown in Eq.\add{~}\eqref{lstmh}. It \add{can be seen}\del{is obvious} from the equations that the output of an LSTM at time step $t$ only depends on $x_t$ and $h_{t-1}$, which are the input at time step $t$ and the information from the previous time steps. Thus, the information from the time steps after $t$ is not used in the equations. \del{Due to this reason}\add{To address this}, we employ\del{ed} two different LSTM structures for \add{the} segmentation module and \add{the} recognition module, which will be covered in \del{}\add{S}ubsection\add{s~\ref{se:SegmentationModule}~and~\ref{se:RecognitionModule}, respectively}\del{ $B$ and $C$}.
Residual connections \cite{resnet} are adopted between adjacent LSTM layers in our model,\del{. B} \add{b}ecause the residual connection encourages the higher layer to learn something different from what the lower layer has already learned. \del{And i}\add{I}t also encourages gradient flow.

\item \add{\it FC Layer}\del{FC Layer}

The FC layers represent a sequence of non-linear transformations to the CNN-LSTM extracted features. Each FC layer corresponds to a linear transformation and a rec\del{i}tified-linear (ReLU) \cite{RELU} activation function. Batch Normalization is \mdel{also} used after each linear transformation and before the activation function, following \cite{BNA}. The FC layer yields different outcomes for different modules. For the segmentation module\del{,} it differentiates if \add{the} current frame belongs to the boundaries of an activity. For the recognition module\del{,} it works as a similarity metric that measures how similar the two inputs are.

\end{itemize}

Following the notation in \cite{beyondTemporalPool}, the shorthand description of each single branch is as follows: $C(64) - C(64) - P - C(64) - C(64) - P - R(128) - R(128)$, where $C(n)$ denotes a convolutional layer with $n$ feature maps, $P$ a max-pooling layer, and $R(n)$ a recurrent LSTM layer with $n$ cells.

\subsection{Segmentation Module}
\add{\label{se:SegmentationModule}}
The periods that blur finishing and beginning two consecutive activities are transitions. \del{It has}\add{They have} features from both sides and contain\del{s} the boundary of activit\del{y}\add{ies}. The exact position of the boundary is \add{often} difficult to define in the data stream. But, we can define that the boundary is in the transition\del{,} because the previous activity has \del{been} finished within this period\mdel{,} and the subsequent activity has \del{been} started within this period. Thus, it is reasonable to \add{either} label the transition \del{either} as an independent \add{transition} activity\mdel{,}\del{: transition, or} \add{or as an unknown activity, or to include it into the} \del{as the} activity before or after \del{itself}\add{the transition period}\del{, or as an unknown activity} \cite{transitionAware}. \add{An u}\del{U}nknown activity represents an infinite space of arbitrary activities. The space of transition\madd{s} is not infinite, but if there are $n$ different activities, there will be $\frac{n!}{(n-2)!}$ kinds of possible transitions that are all very short. These reasons make learning a\add{n explicit} model for transition and unknown activity difficult. However, transitions and unknown activities occur between activities of interest\del{,}\add{;} this characteristic \del{makes}\add{allows} transition and unknown activity \del{can}\add{to} be used as the boundary period to the activity sequence. It is important to notice that if the transition and unknown activity are used as the boundary of an activity, then the segmentation is not a hard segmentation\del{,} \add{but} rather a soft segmentation. The position of a soft segmentation is not a single frame\del{,} but a period that contains one or more than one frame\del{s}. \add{To address this ambiguity of the transition period, multiple segmentations and classifications of the resulting segments should be considered accurate representations of the data.}

Table \ref{tab:seg_error_table} shows the error assessment \cite{transitionAware} used in this paper for \add{judging correct} segmentation and recognition. In the table, A, B, C are activities of interest, U is unknown activity, T is transition.
\begin{table}[h!]
  \begin{center}
    \caption{Error assessment table.}
    \label{tab:seg_error_table}
    \begin{tabular}{l|c|r} 
      \textbf{Ground-Truth} & \textbf{Segmentation \& Recognition} & \textbf{Error Evaluation}\\
      \hline
      \multicolumn{3}{c}{Basic Activities} \\
      \hline
      A-A-A & A-A-A & correct\\
      A-A-A & A-B-A & incorrect\\
      A-A-A & A-T-A & incorrect\\
      A-A-A & A-U-A & incorrect\\
      \hline
      \multicolumn{3}{c}{Without Transitions} \\
      \hline
      A-B & A-B & correct\\
      A-B & A-C-B & incorrect\\
      A-B & A-T-B & incorrect\\
      A-B & A-U-B & incorrect\\
      \hline
      \multicolumn{3}{c}{With Transitions} \\
      \hline
      A-T-B & A-B & correct\\
      A-T-B & A-C-B& incorrect\\
      A-T-B & A-T-B & correct\\
      A-T-B & A-U-B & correct\\
    \end{tabular}
  \end{center}
\end{table}

Segmenting a data sequence with the proposed \mdel{segmentation} module can be formulated as \add{follows}: given \add{the} current time step $t_{M}$\del{,} \add{and} an activity sequence $S = <a_0, \ldots, a_M, \ldots, a_N>$ from time step $t_0$ to $t_N$, $(0<M<N)$, the proposed segmentation module \del{differentiates}\add{predicts whether} \del{if} $t_{M}$ is a segmentation frame in the data sequence. We consider the boundary as a frame or a list of consecutive frames that \add{indicate that} the activities before and after it are different. Thus, we split the sequence $S$ at time step $t_M$ into a pair of subsequences $(sub_A, sub_B)$.
Let $sub_A = <a_0, \ldots, a_M >$ be the activity sequence from time step $t_0$ to $t_M$, \mdel{namely,}\madd{representing} the history \mdel{sequence} \madd{of}\mdel{that contains} information from previous time steps. And let $sub_B = <a_N, \ldots, a_{M+1}>$ be the activity sequence from time step $t_{M+1}$ to $t_N$ in reverse chronological order, \mdel{namely,}\madd{representing} the future sequence that contains information from a future time step to \add{the} current time step. We restrict the length of the future sequence, so that the time delay in the system is not infinite.

As shown in Fig. \ref{siameseArch}, for the segmentation module\del{,}
$Branch_{A}$ is responsible to learn a \com{I am not a really big fan of these really long and descriptive names as symbols. how about giving it a name ? E.g.  "Representation Vector $\vec{V}_{history}$" or something similar ?}
representation vector $\vec{V}_{history}$ from \add{the} history sequence, while $Branch_{B}$ is responsible to learn a representation vector $\vec{V}_{future}$ from the future sequence. 
We assume \add{here that} if $t_{M}$ is a segmentation frame, $\vec{V}_{history}$ and $\vec{V}_{future}$ should be far \add{a}way \add{from each other} in the representation space\mdel{,}\madd{;} otherwise \add{they} should be close \add{as they are then considered to be part of the same activity}. \del{And this}\add{In this way, the} distance metric \add{encoded in the representation space} represents the transition process nicely, because the distance between the two vectors changes from short to long when \add{the} previous activity is finishing and \add{the} subsequent activity is starting.

The two branches in the segmentation siamese network use the information before and after \add{the} current time step explicitly and process the data sequence in opposite directions. \del{But t}\add{T}he LSTM employed in each branch is a uni-directional LSTM, not a bi-directional LSTM (BLSTM). \add{The reason for this is that} \cite{LSTMAE} shows that\del{,} in an LSTM autoencoder\del{,} \add{where} the decoder decodes the target sequence in reverse order\add{, it} can help the model \add{to capture}\del{catch} the short range correlations easily. Thus, $Branch_{A}$ process\mdel{es}\madd{ing} the history data in chronological order while $Branch_{B}$ process\add{es}\del{ing} the future data in reverse chronological order also aims to \del{catch}\add{capture} short range correlations\del{,} because short range correlations can restrict the segmentation in a short region, while long range correlations can\del{'t} \add{not}.

The FC layers after the LSTM learn a non-linear distance metric to measure the similarity between $\vec{V}_{history}$ and $\vec{V}_{future}$, then output this distance. A generalized Gaussian function\del{:}\add{,} $\mathcal{G}\mathcal{G}(x; \beta) = \frac{\beta}{2\alpha\Gamma(1/\beta)}e^{-(|x-\mu|/\alpha)^{\beta}}$\add{,} is used as the distance target to train the model. $\beta$ here is \add{a} shape parameter of the function\del{,}\add{;} if $\beta$ is picked properly, the generalized Gaussian function can have a flat top and sharp edge that represent\add{s} the occurrence and duration of a transition between activities. In the test stage, we use the segmentation module as a classifier. For an input frame, if the output from the segmentation module is above a threshold, we consider the input frame as a segmentation frame. Note that, since we adopt soft segmentation, it is possible that a list of consecutive frames can all be \add{identified}\del{differentiated} as segmentation positions.

\subsection{Recognition Module}
\add{\label{se:RecognitionModule}}

The recognition module sticks to the same siamese architecture described in \del{s}\add{S}ubsection\add{~\ref{se:Siamese}}\del{ $A$}, but the LSTM layers are different. \add{In contrast to the segmentation network, a}\del{A} BLSTM is adopted here, which use\add{s} two uni-directional LSTMs working on the same data but in opposite directions along the time domain. \del{And t}\add{T}he outputs from both uni-directional LSTMs are concatenated to be fed into the next layer of the network. The BLSTM structure aims to extract the information from the full sequence\mdel{,} and balance the importance of both sides of the sequence. In comparison to the LSTM layers used in the segmentation module, the BLSTM used here learns to capture both short and long range correlations in the data sequence. 

The recognition siamese network is trained with triplets $(x_A, x_B, y) $, where $x_A$ and $x_B$ are the segmented sequences and $y \in  \left\{0, 1\right\}$ indicates that $x_A$ and $x_B$ are the same kind of activity,\del{ if} $y = 1$\add{,} or different kinds of activit\mdel{y}\madd{ies}, \del{if} $y = 0$.

The similarity function for the recognition siamese network is predefined as\del{:} $D(x_A,x_B)=exp(-|f(x_A)-f(x_B)|) \in \left\{0,1\right\}$ \cite{sentenceSiamese}. It forces the model to learn a mapping $f(x) $ that captures the critical similarities between the input pairs $(x_A, x_B) $, and if $y = 1$, then $f(x_A), f(x_B) \in R^d$ should stay close, while $f(x_A), f(x_B)$ should be far away \add{from each other} if $y = 0$. The mean-squared-error (MSE) is used to measure the loss between the model estimated similarity and the ground-truth similarity.

Assuming a list of segmented sequences: $l = \left\{act\_seq_0, \ldots, act\_seq_M, \ldots, act\_seq_N\right\}$, the trained recognition siamese network maps all the activity segments in $l$ into the representation space. Then, different clustering algorithms can be used on those representation vectors to build the clusters. Here, single-linkage clustering \add{is adopted}. A drawback of single-linkage is that it tends to build long thin clusters in which the nearby elements of the same cluster have small distance values\del{, however,} \add{while} the elements at the edge of a cluster may have larger distance to the elements from the opposite side of the same cluster\del{, but may have smaller distance} \add{than they have} to the elements of other clusters \cite{clusteranalysis}. This may be a problem in other research areas, but it aligns with some features of human activity sequence\add{s}. \add{Here, t}\del{T}he data stream of activities consists of activities of interest interwoven with transitions. As discussed above, the exact borders of activities and transitions are difficult to define. The beginning and ending stages of a transition can be viewed as part of the previous and subsequent activities that it connects with. \add{However, it should be noticed}\del{But notice} that transitions between different combinations of activities are essentially different, e.g. the transition from $lying$ to $sitting$ can not be the same as the transition from $walking$ to $standing$. These transitions may have a large distance between each other, but are close to their connected activities\add{,} respectively. The single-linkage clustering preserves this feature naturally.

\section{EVALUATION AND EXPERIMENT}

\subsection{Datasets}
For empirical evaluation and comparison with other approaches, we test our method on three public datasets that contain the raw sensor data sequence of human activity. The descriptions of the datasets are listed below\del{:}\add{.}

The \textbf{Daphnet Gait dataset (DG)} \cite{dg}: This dataset is collected from 10 patients with Parkinson's disease. Three triaxial acceleration sensors are fixed at the patient's ankle, upper leg, and trunk to record the activities performed by the patient with \del{the}\add{a} sampling rate of 64Hz. The patients were instructed to carry out activities that are likely to induce freezing of gait (FoG). The FoG is a common symptom in Parkinson's disease that affects patient's activities like $walking$. The objective is to identify the FoG of the patients. We follow the same settings as in \cite{dnnHARbenchmark}\del{,} \add{and} use the records of patient 9 for validation, the records of patient 2 for test, and the rest for training.

The \textbf{WISDM dataset} \cite{wisdm}: This dataset contains data collected from 36 users perform\add{ing}\del{ed} 6 different activities, \del{e.g.}\add{including} $jogging$, $ascending$ $stairs$, \add{etc., }in a controlled experiment environment. The data is recorded with one triaxial accelerometer in a smartphone at a sampling rate \add{of} 20Hz.

The \textbf{SBHAR dataset} \cite{sbhar}: This dataset provides data from a group of 30 volunteers with \add{an} age bracket of 19-48 years, carrying out 6 basic activities, such as $walking$\del{,} \add{and} $lying$, and 6 postural transitions such as $stand-sit$\del{,} \add{and} $sit-lie$. The data is recorded by letting the volunteers wear a smartphone on the waist during the experiment\del{ execution}. The smartphone's embedded triaxial accelerometer and a gyroscope recorded the data at a sampling rate of 50Hz.

\subsection{Performance Measures}
In the experiment\add{s}, we use two performance metrics: many-to-one accuracy and weighted $F_1$ score.

Because our model learned a metric without explicit labels, \del{and} there is no straightforward way to compare the ground truth labels with \add{the} clusters found by \del{the}\add{an} unsupervised algorithm using our learned distance metric. Here, we use \madd{many-to-one accuracy as a} \mdel{the} mapping-based performance measures\mdel{: many-to-one accuracy} to evaluate the performance. For the cluster with label $t$, select the most frequent correct label $t^*$ for those frames in the cluster. Replace $t$ with $t^*$. After processing each cluster this way, compute the accuracy as usual. 

The weighted $F_1$ score has been used as the performance metric (for the DG dataset) in related work. In order to compare our results to the state-of-the-art we also calculate the weighted $F_1$ score:

\begin{align}
&F_w = 2\sum_{i=1}^{C} \frac{N_i}{N_{total}} \frac{Precision_i \times Recall_i}{Precision_i + Recall_i} 
\end{align}
where $N_i$ is the number of samples in class $i$, $N_{total}$ is the total number of samples, and for the given class $i$, $Precision_i = \frac{TP_i}{TP_i + FP_i}$, $Recall_i = \frac{TP_i}{TP_i + FN_i}$ and $i = 1, \ldots, C$ is the set of classes. $TP_i$, $FP_i$ represent the number of true positive\add{s} and false positive\add{s}, $FN_i$ is the number of false negative\add{s}.

\subsection{Experimental Results}
Table\add{s} \ref{tableDG}, \ref{tableWISDM}, and \ref{tableSBHAR} illustrate the comparison of the proposed method against existing supervised and unsupervised methods on \add{the} DG, WISDM and SBHAR datasets.

\begin{table}[ht]
\caption{Results on \add{the} DG dataset in terms of weighted $F_1$ score.}
\begin{center}
\begin{tabular}{l|c}
\textbf{Method} & \textbf{$F_{1}$ Score}\\

    \hline
    LSTM baseline \cite{understandRNNHAR} &0.6675\\ 
    DeepConvLSTM \cite{understandRNNHAR} &0.7344\\ 
    LSTM-S \cite{dnnHARbenchmark} & 0.7600 \\
    LSTM + Continuous Temporal, Sensor Attention \cite{understandRNNHAR}  &0.8373\\
    Proposed Method & \textbf{0.8952} \\
    
\end{tabular}
\end{center}
\label{tableDG}
\end{table}
 
\begin{table}[ht]
\caption{Results on \add{the} WISDM dataset in terms of accuracy.}
\begin{center}
\begin{tabular}{l|c}
\textbf{Method} & \textbf{Accuracy(\%)}\\
 
    \hline
    Multilayer Perceptron \cite{wisdm} &91.7\\ 
    Ensemble Learning \cite{ensembleResult}  & 94.3\\
    CNN with partial weight sharing \cite{CNNmobicase}  & 96.88 \\
    DBN+HMM\cite{DBNHAR}  & \textbf{98.23} \\
    Proposed Method & 96.68 \\
        
\end{tabular}
\end{center}
\label{tableWISDM}
\end{table}

The results in \del{t}\add{T}able \ref{tableDG} show\del{s} that the proposed weakly supervised method outperforms the state-of-the-art, LSTM with continuous temporal and sensor attention \add{\cite{understandRNNHAR} }, on \add{the} DG dataset. Table\add{s} \ref{tableWISDM}, \ref{tableSBHAR} show that the best performance on \add{the} WISDM and SBHAR datasets are achieved by supervised methods, DBN$+$HMM \add{\cite{DBNHAR}}  and Convolutional Neural Network \add{\cite{lundHAR}}, \add{respectively\mdel{,}}\madd{. H}\mdel{h}owever\madd{,}\del{,} expensive labeled data is required to train these models. \begin{table}[h!t]
\caption{Results on \add{the} SBHAR dataset in terms of accuracy.}
\begin{center}
\begin{tabular}{l|c}
\textbf{Method} & \textbf{Accuracy(\%)}\\
    \hline
     Probability SVM with Filter \cite{transitionAware} & 96.78\\
     Convolutional Neural Network \cite{lundHAR} & \textbf{98.7}\\
     CNN Autoencoder with K-means \cite{lundHAR} &55.0\\
     PCA with K-means \cite{lundHAR} &62.1  \\
     Proposed Method & 92.68 \\
 
\end{tabular}
\end{center}
\label{tableSBHAR}
\end{table} 
Although \del{the}\add{completely} unsupervised methods do not need any labeled data, their performance are not comparable to supervised methods \add{\cite{lundHAR}}. The proposed weakly supervised method is trained under limited supervision by using only the information about the similarity between the activities\del{, while it outperforms unsupervised methods} and achieves comparable results to supervised methods on \add{the} WISDM and SBHAR datasets \add{while significantly outperforming the unsupervised methods}.

\com{The legends in these figures are way too small. Can we make them bigger ?}
\subsection{Visualization of the Representation Space}
\begin{figure*}
    \centering
    \begin{subfigure}[b]{0.329\linewidth}        
        \frame{\includegraphics[width=0.95\textwidth]{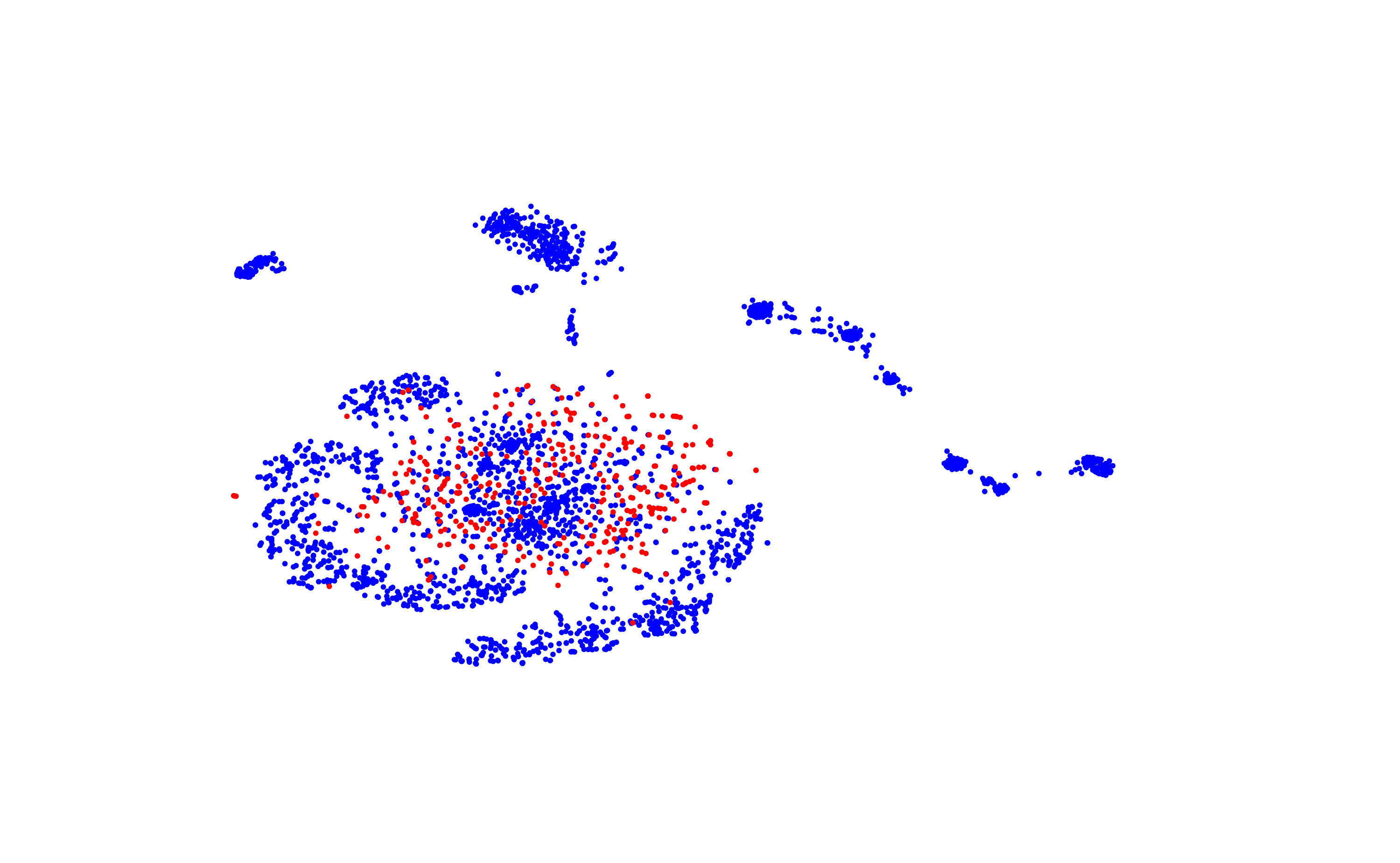}}
        \caption{Original Input Space}
        \label{oridfog}
    \end{subfigure}
    \begin{subfigure}[b]{0.329\linewidth}        
        \frame{\includegraphics[width=0.95\textwidth]{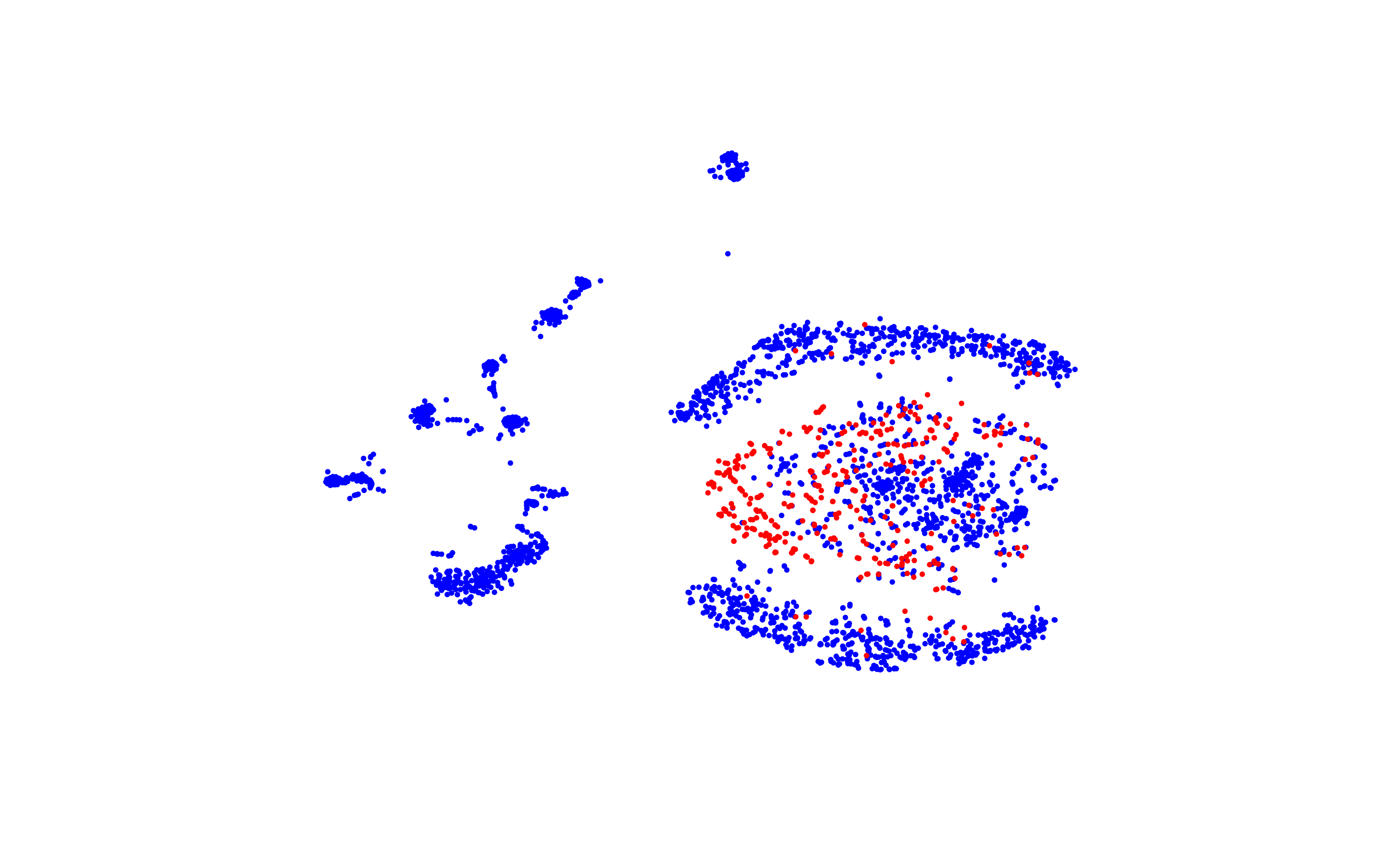}}
        \caption{CNN Autoencoder}
        \label{cnnAEdfog}
    \end{subfigure}
     \begin{subfigure}[b]{0.329\linewidth}        
        \frame{\includegraphics[width=0.95\textwidth]{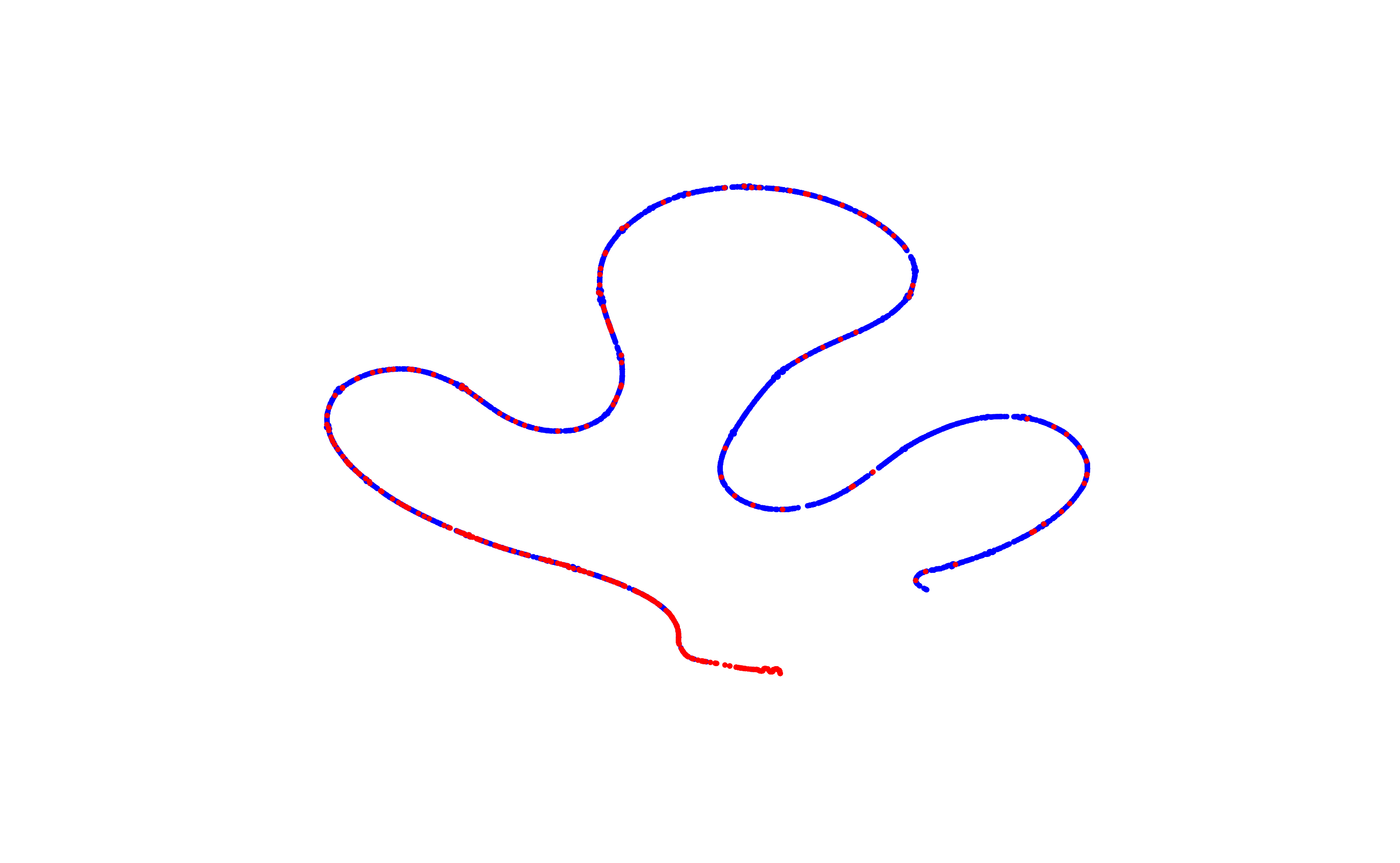}}
        \caption{Proposed Method}
        \label{myModelHiddfog}
    \end{subfigure}
    \caption{t-SNE visualizations on \add{the} DG dataset: 
    \legendsquare{fill=blue}~Freezing of gait,
    \legendsquare{fill=red}~No freezing.
    \protect\subref{oridfog} Original input space, \protect\subref{cnnAEdfog} Representation space from \add{the} cnn autoencoder and, \protect\subref{myModelHiddfog} Representation space from the proposed method.}\label{dfog}
\end{figure*}

\begin{figure*}
    \centering
    \begin{subfigure}[b]{0.329\linewidth}        
        \frame{\includegraphics[width=0.95\textwidth]{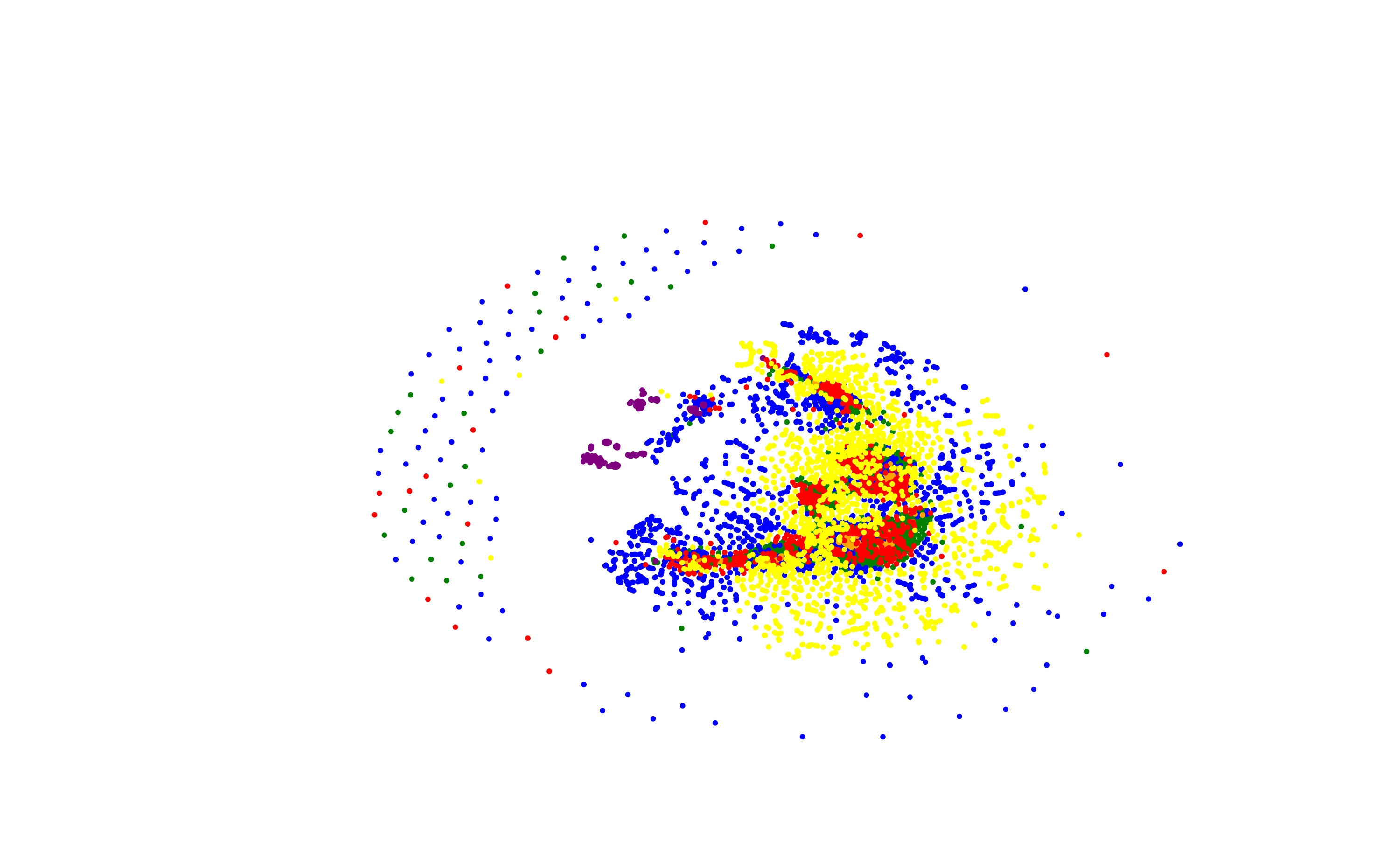}}
        \caption{Original Input Space}
        \label{origiwisdm}
    \end{subfigure}
    \begin{subfigure}[b]{0.329\linewidth}        
        \frame{\includegraphics[width=0.95\textwidth]{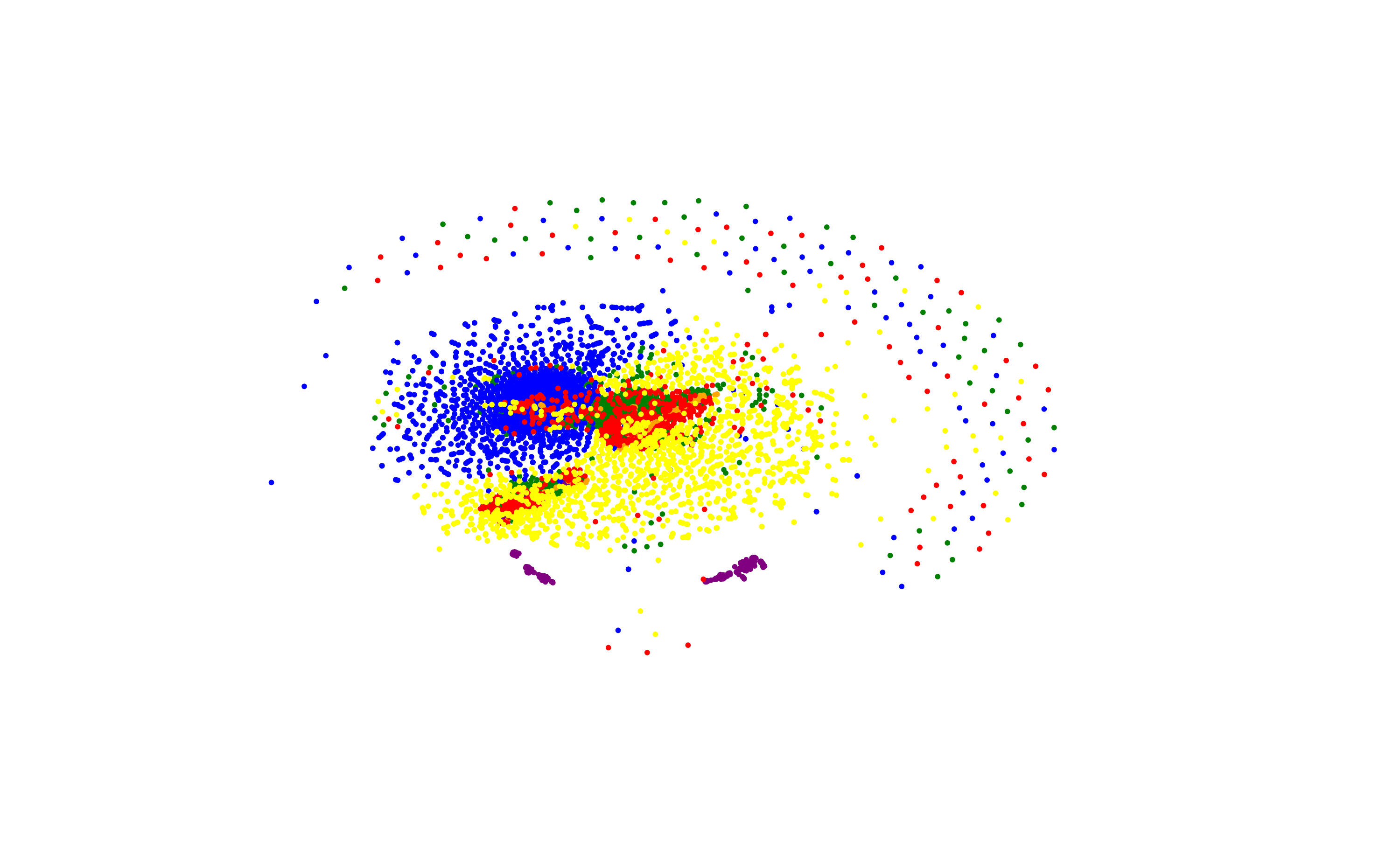}}
        \caption{CNN Autoencoder}
        \label{cnnAEwisdm}
    \end{subfigure}
     \begin{subfigure}[b]{0.329\linewidth}        
        \frame{\includegraphics[width=0.95\textwidth]{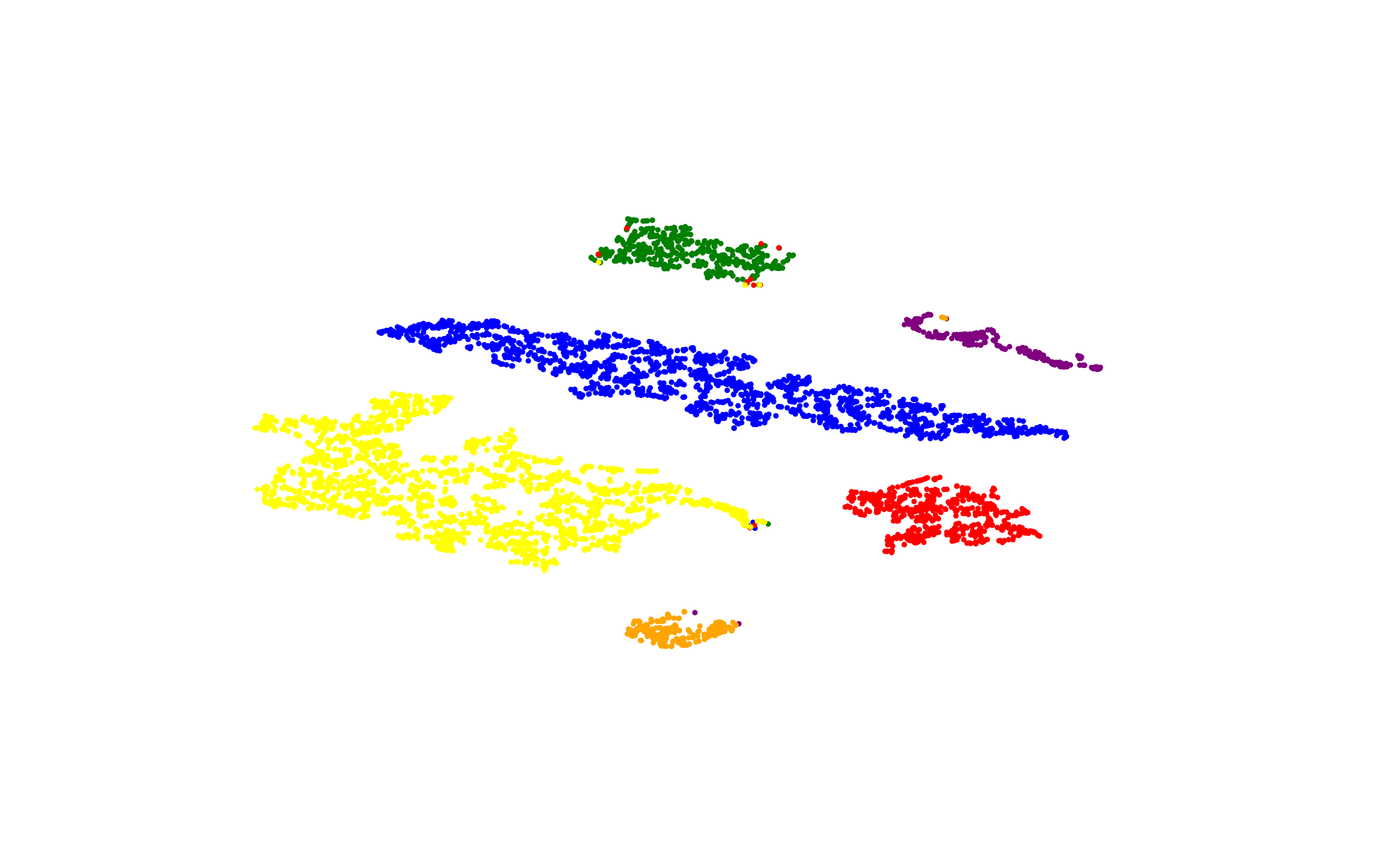}}
        \caption{Proposed Method}
        \label{myModelHidwisdm}
    \end{subfigure}
    \caption{t-SNE visualizations on \add{the} WISDM dataset: 
     \legendsquare{fill=blue}~Jogging,
    \legendsquare{fill=Plum}~Sitting,
    \legendsquare{fill=YellowOrange}~Standing,
    \legendsquare{fill=Yellow}~Walking,
    \legendsquare{fill=red}~Ascending stairs,
    \legendsquare{fill=Green}~Descending stairs.
    \protect\subref{origiwisdm} Original input space, \protect\subref{cnnAEwisdm} Representation space from \add{the} cnn autoencoder and, \protect\subref{myModelHidwisdm} Representation space from the proposed method.}\label{WISDM}
\end{figure*}

\begin{figure*}
    \centering
    \begin{subfigure}[b]{0.329\linewidth}        
        \frame{\includegraphics[width=0.95\textwidth]{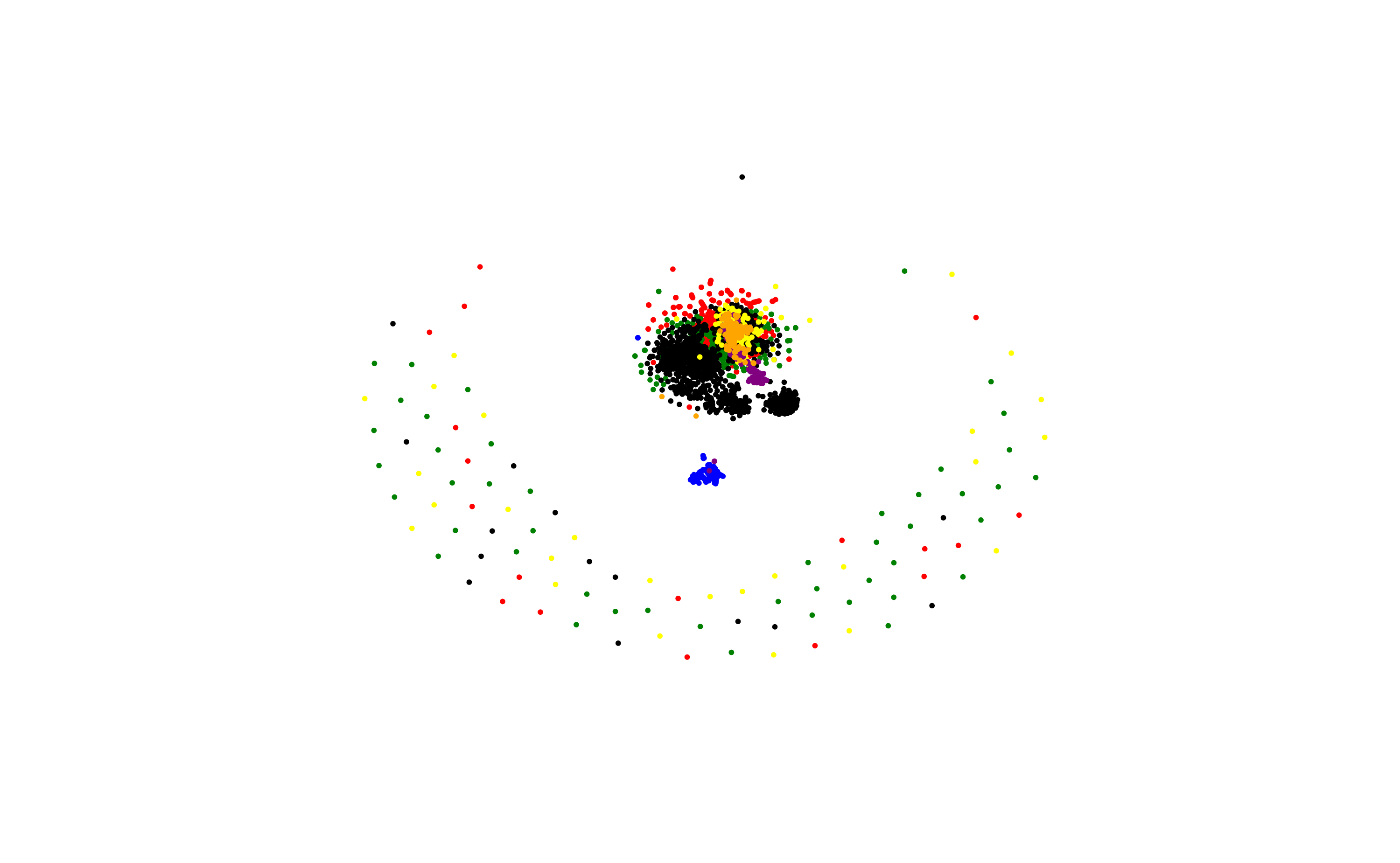}}
        \caption{Original Input Space}
        \label{origiSBHAR}
    \end{subfigure}
    \begin{subfigure}[b]{0.329\linewidth}        
        \frame{\includegraphics[width=0.95\textwidth]{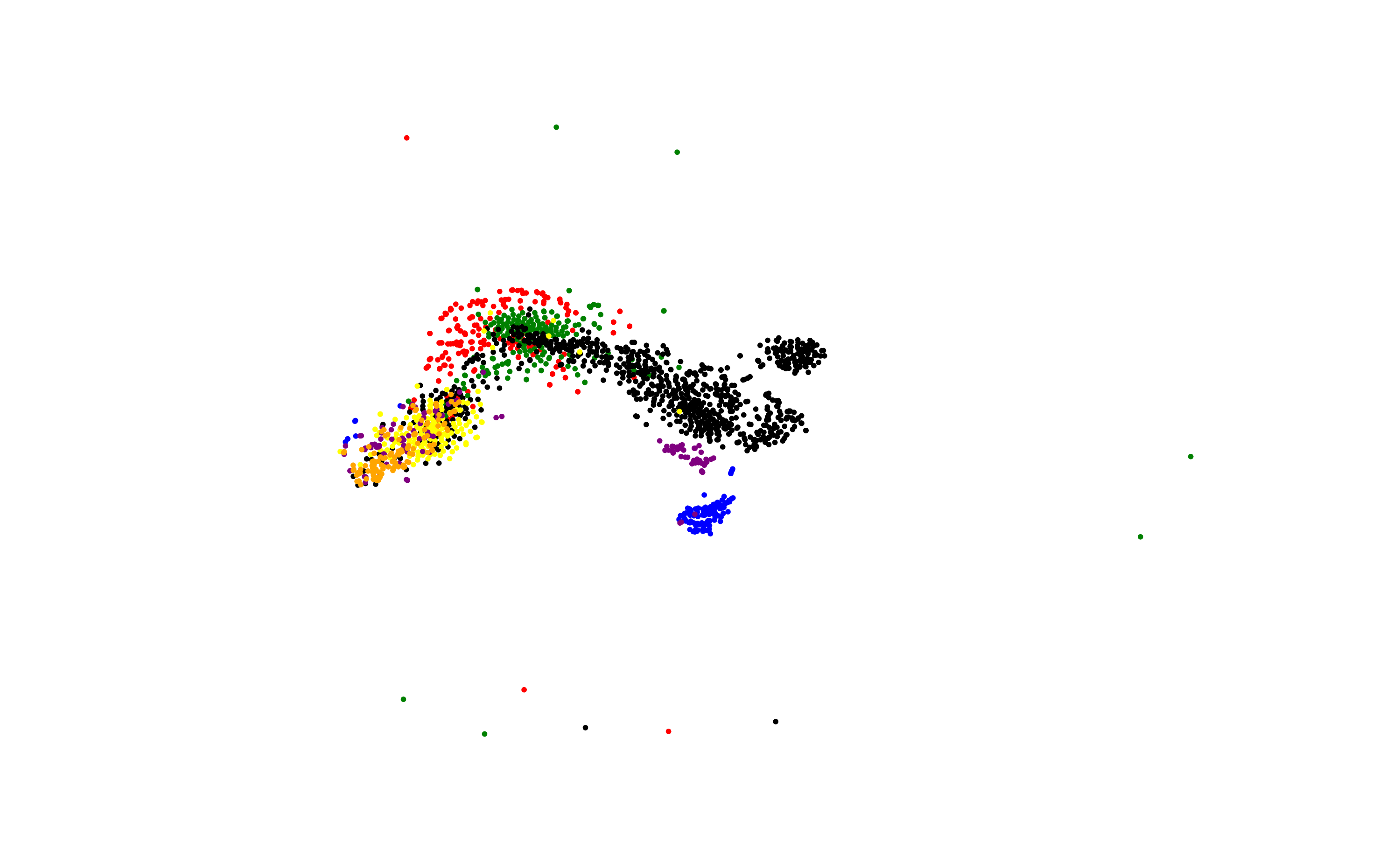}}
        \caption{CNN Autoencoder}
        \label{cnnAESBHAR}
    \end{subfigure}
    \begin{subfigure}[b]{0.329\linewidth}        
        \frame{\includegraphics[width=0.95\textwidth]{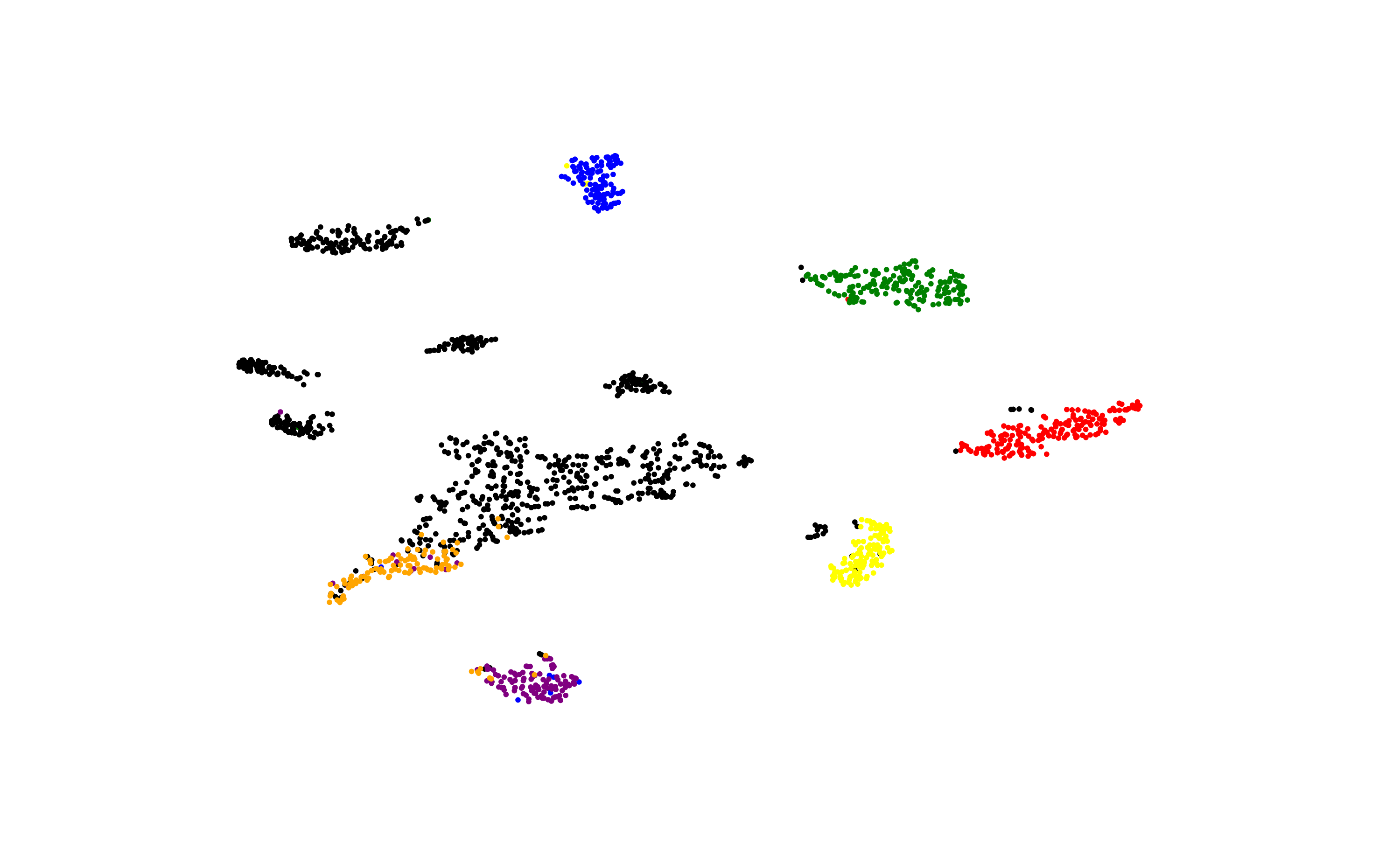}}
        \caption{Proposed Method}
        \label{myModelHidSBHAR}
    \end{subfigure}
    \caption{
     t-SNE visualizations on \add{the} SBHAR dataset:
    \legendsquare{fill=blue}~Lying,
    \legendsquare{fill=Plum}~Sitting,
    \legendsquare{fill=YellowOrange}~Standing,
    \legendsquare{fill=Yellow}~Walking,
    \legendsquare{fill=red}~Walking upstairs,
    \legendsquare{fill=Green}~Walking downstairs,
    \legendsquare{fill=Black}~Transitions \& Unknown activities.
    \protect\subref{origiSBHAR} Original input space, \protect\subref{cnnAESBHAR} Representation space from \add{the} cnn autoencoder and, \protect\subref{myModelHidSBHAR} Representation space from the proposed method. 
    }\label{SBHAR}
\end{figure*}

In order to better understand the distribution of the activity vectors in the representation space, we applied t-sne \cite{tsne} to map the representation vectors to two dimensions for visualization. \del{See}\add{The results are shown in} Figures \ref{dfog}, \ref{WISDM}, and \ref{SBHAR} for dataset DG, WISDM and SBHAR\add{,} respectively.

Fig. \ref{oridfog}, \ref{origiwisdm}, and \ref{origiSBHAR} show the activity sequences in the original input space. In DG and WISDM all the activities are mixed together. In SBHAR only the activity $lying$ is separable from other activities. 

Fig. \ref{cnnAEdfog}, \ref{cnnAEwisdm}, and \ref{cnnAESBHAR} show the results produced by a temporal convolutional autoencoder. The idea is to let the autoencoder compress the input activity sequence into a representation vector and reconstruct the sequence from this representation. It is expected that this compressed representation contains key information of the input sequence that is useful to reconstruct itself and to differentiate it from other activities. The results show that after being mapped into the representation space by the autoencoder, some representation vectors are almost separable. In SBHAR the vectors of $lying$, $sitting$ and $standing$ are separated from other activities. In WISDM the cluster of $sitting$ is well formed. Thus, it is clear that the separation between static posture activities $\left\{lying, sitting, standing\right\}$ and dynamic activities $\left\{jogging, walking, ascending \: stairs, descending \: stairs\right\}$ is almost solved by the mapping of \add{the} autoencoder. It confirms that \add{the} autoencoder can capture useful information of the activities, especially the difference between static and dynamic activities. However, the mapping of \add{the} autoencoder is not powerful enough to further group each activity into its own cluster and \add{it}\del{autoencoder} is not able to capture the difference between FoG and normal activities, so in \add{the} DG dataset, the FoGs are still mixed with normal activities.

Fig. \ref{myModelHiddfog}, \ref{myModelHidwisdm}, and \ref{myModelHidSBHAR} are the results generated by \add{the} proposed method. In DG a long thin cluster is built. The possible reason is that most of the normal activities in the experiment are normal $walking$, and no significant difference exists between different $walking$ sequences. FoG is an abnormal activity \add{that} shows up during normal $walking$, thus FoG also carries the characteristics of normal $walking$. Due to this reason, a long thin cluster can represent the process of normal $walking$ \add{that} transfers to FoG gradually. As shown in Fig. \ref{myModelHiddfog}, FoGs stay on one side of the cluster, while normal $walking$ activities stay on the other side. In WISDM and SBHAR each kind of activity is grouped into its own cluster which is clearly separable. Note that\mdel{,} in \del{f}\add{F}ig. \ref{myModelHidSBHAR}\del{,} the transitions and unknown activities are viewed as an independent special activity and grouped into 6 clusters. This result confirms our assumption in Subsection \ref{se:RecognitionModule} that transitions between different kinds of activities are essentially different, and the unknown activities represent an infinite space of arbitrary activities. Although we treat all the transitions and unknown activities the same way, \del{but} a long distance for different kinds of transitions and a small distance within the same kind of transition\mdel{s} \del{are}\add{is} still maintained.

\section{CONCLUSIONS}

In this paper, we proposed a weakly supervised method that can learn to segment and recognize human activit\mdel{y}\madd{ies} under limited supervision without using \del{the} explicit labels. \add{The proposed method uses a \mdel{S}\madd{s}egmentation and a \mdel{R}\madd{r}ecognition module built around a common \mdel{S}\madd{s}iamese network architecture consisting of CNN and LSTM layers to capture temporal relations efficiently.} We verified its effectiveness on three HAR datasets and further analyzed the learned mapping function by visualizing the activity vectors in the representation space. The result\add{s} \del{proves}\add{show} that the proposed model can map the activity sequences into a space where the distance metric indicates the similarity of the activities. The learned distance metric can be applied with different clustering algorithms and achieve state-of-the-art or comparable performance to supervised method\add{s}.

\addtolength{\textheight}{-12cm}   




\bibliographystyle{ieeetr}
\bibliography{smc2019.bib} 

\end{document}